\documentclass[prl,twocolumn,amsmath,a4paper]{revtex4}

\usepackage{graphicx}

\begin{document}
\title{
Conductance correlations in a mesoscopic spin glass wire : \\
 a numerical Landauer study}

\author{Guillaume Paulin}
\affiliation{CNRS - Laboratoire de Physique de l'Ecole Normale
Sup{\'e}rieure in Lyon, \\
46, All{\'e}e d'Italie, 69007 Lyon, France}

\author{David Carpentier}
\affiliation{CNRS - Laboratoire de Physique de l'Ecole Normale
Sup{\'e}rieure in Lyon, \\
46, All{\'e}e d'Italie, 69007 Lyon, France}

\date{\today}

\begin{abstract}
In this letter we study the coherent electronic transport through a metallic nanowire with 
magnetic impurities. The spins of these impurities are considered as frozen to mimic a low temperature 
spin glass phase. The transport properties of the wire are derived from a 
numerical  Landauer  technique which provides the conductance of the wire as a function of the disorder 
configuration. We show that the correlation of conductance between two spin configurations provides a measure 
of the correlation between  these spin configurations. This correlation corresponds to the mean field
 overlap in the absence of any spatial order between the spin configurations. 
 Moreover, we find that these conductance correlations are sensitive to the spatial order between 
 the two spin configurations, {\it i.e} whether the spin flips between them occur in a compact region 
 or not. 
 \end{abstract}

\maketitle
Spin glasses have been a focus of continuous interest in condensed matter for more than three decades. 
In spite of the relative simplicity of the models describing their physics, a precise understanding of their 
properties  remains elusive. Spectacular progress has been made in  understanding their 
nature at the mean field level \cite{Parisi:1980,Mezard:1987}, in characterizing their exotic aging properties in mean field models \cite{Cugliandolo:2002}, 
including a proper description of the violation of the fluctuation-dissipation theorem, and experimentally 
in characterizing their memory and rejuvenation effects (see \cite{Vincent:2008} for a recent review). 
However the applicability of mean field ideas in 
real samples remains debated \cite{Fisher:1986}, with alternative approaches stressing the  importance of the nature of 
excitations, and their consequences on various out-of-equilibrium properties of the 
phase\cite{Houdayer:2000}.

A crucial quantity to characterize this spin glass physics is the correlation between 
different states of spins $\{\vec{S}_i^{(1)}\}_{i}$ and $\{\vec{S}_i^{(2)}\}_{i}$  in a given sample corresponding 
{\it e.g } Êto two different times $t_{1}$ and $t_{2}$  in a same quench, or two different quenches. 
 For a single spin $i$, this correlation  is naturally given by the local overlap $\vec{S}_i^{(1)}.\vec{S}_i^{(2)}$. For 
 a collection of spins, mean-field theory neglects any spatial correlation of this local overlap : the correlation 
 between the two spin states is given by 
\begin{equation}
Q_{12} = \frac{1}{N}\sum_{i=1}^N\vec{S}_i^{(1)}.\vec{S}_i^{(2)}. 
\label{equ:overlap}
\end{equation}
 The distribution of this overlap between states reached after successive cooling in a sample plays a central 
 role in the Parisi's mean field theory. 
Note however that this overlap (\ref{equ:overlap}), while 
perfectly adequate at the mean field level, does not contain any information on the geometry 
of the correlation. In the simplest case of Ising spins it  simply counts the number of spin flips between the two 
spin states, without any information on  whether these spin flips occur  in a compact region or randomly in the 
sample. Information about the spatial structure of this spin states correlation would require a more refined function. 

 Recently, building on previous theoretical work on sensitivity of conductance fluctuations to perturbations like 
magnetic impurities \cite{Altshuler:1985} and pioneering experiments on conductance fluctuations in spin glasses 
\cite{deVegvar:1991,Jaroszynski:1998,Neuttiens:2000}, the study of magneto-conductance of spin glass nanowires was proposed 
as a unique probe of these correlations between spin glass configurations \cite{Carpentier:2008}. Indeed, the 
correlation between conductances for two different mean-field like spin states depends monotonously on the overlap between 
these two states. Hence measurement of this conductance correlation can give access to the corresponding overlap 
\cite{Carpentier:2008,Carpentier:2009}. 
 This proposal calls for experimental and numerical studies of the correlations of conductance in a spin glass metallic 
 system.
  It is the purpose of this letter to develop a numerical study of these conductance correlations, 
  and in particular to  
 address the question of sensitivity of these conductance correlations 
 to spatial order between the corresponding spin states, originating from {\it e.g} the nature of excitation in the 
 spin glass (see \cite{Cieplak:1991} for an alternative numerical approach focused on the time evolution of conductance fluctuations).  This question is naturally of crucial importance for 
 experimental studies of quantum transport in spin glass nanowires.   
 To address this question,  we present a numerical Landauer approach allowing to accurately describe the weak 
localization regime of experimental relevance. This approach allows to go beyond the restrictions of 
analytical techniques and consider random spin states with spatial correlations between them. 
 
   
  To describe the electronic transport in the low temperature phase of a spin glass metallic wire, we consider a 
 tight-binding Anderson model with magnetic disorder : 
 \begin{multline}
\mathcal{H} = \sum_{<i,j>,s}t_{ij}c_{j,s}^{\dagger}c_{i,s} + \sum_{i,s}v_ic_{i,s}^{\dagger}c_{i,s}  \\
 +  J\sum_{i,s,s'}{\vec{S}}_i .  {\vec{\sigma}}_{s,s'}c_{i,s}^{\dagger}c_{i,s'},
\end{multline}
 where $t_{ij}=t$ represents the hopping of an electron from site $i$ to $j$, $v_i$ is the scalar random 
 potential uniformly distributed in the interval $[-W/2;W/2]$ and $J$ is the intensity of the magnetic disorder
  which is typically smaller than $W$. The $\vec{\sigma}$  are Pauli matrices and 
 $s$ labels the spin state of the electron. 
 The magnetic impurities of the 
 spin glass contribute to two different random potentials: a scalar diffusive potential $v_{i}$ 
 (originating in part from 
 the random positions of the impurities) and a magnetic disorder originating from the random 
 spins ${\vec{S}}_i$.  
 In this description, all impurity spins $\vec{S}_{i}$ are frozen 
  and treated as classical spins. 
 We choose the classical spins 
 $\vec{S}_i$ randomly in the sphere of radius $S$, and independent from each other. 
 This amounts to  neglect any spatial order in a given state, in agreement 
 with neutron scattering experiments \cite{Mydosh:1993}. Going beyond this simple description by 
 including more complex hidden spatial order goes beyond the scope of the present paper. 
 Owing to the experimental findings of  universal conductance fluctuations  in the spin glass phase\cite{deVegvar:1991,Jaroszynski:1998},  we focus on the corresponding regime where  the wire's length $L_{x}$ is comparable or smaller than the 
  inelastic dephasing length $L_{\phi}$, which effectively includes contribution from free spins.  
  Without loss of generality, we will restrict ourselves to a two-dimensional ribbon of size (in units of lattice spacing) 
 $L_{x}\times L_{y}$ with $L_{y}\ll L_{x}$. 
 
 For a given configuration of scalar disorder $V\equiv \{v_{i}\}_{i}$ and spins $\{{\vec{S}}_{i}\}_{i}$, 
 we numerically determine the corresponding dimensionless conductance $g = G\times h/e^2$  
  through the Landauer formula \cite{Landauer:1957}:
 $ g = \sum_{n,m}\left| t_{nm} \right|^2$, 
where $n$ (resp. $m$) labels the propagating modes in the contacts and $t_{nm}$ the corresponding 
transmission amplitude.  These transmission amplitudes are deduced from the electron's retarded Green's
functions $G^R$ using the Fisher-Lee relation \cite{Fisher:1981}. 
 This Green's function $G^R$ for the system connected to two semi-infinite leads is obtained 
 by a recursive method \cite{MacKinnon:1980}. 
 The Fermi Energy is chosen so that the total number of transverse propagating modes is equal to  
 $2\times L_y$. In units of  $t=1$, the amplitude of scalar disorder is 
 chosen as $W=0.6$ , while the coupling $J$ is varied from 0 (no magnetic disorder) to 0.4 
 ("strong" magnetic disorder). 
 For fixed parameters, the conductance $g$ is a random function of both disorders
  $V$ and $\{{\vec{S}}_{i}\}_{i}$. 
We focus on the weak localization regime, where the conductance $g$ displays universal fluctuations 
of order $1$. 
Experimentally, these fluctuations are measured as a function of a weak  transverse magnetic flux, 
assuming the ergodic hypothesis (see \cite{Tsyplyatyev:2003}  and \cite{Paulin:2009b} for a numerical analysis). 
The amplitude of magneto-conductance fluctuations in a spin glass sample is given by the variance
of the distribution of $g[V,\{{\vec{S}}_{i}\}_{i}]$ as $V$ is varied. 
In the rest of the article, for each configuration of spins  the corresponding distribution will be sampled by $5000$ independent realizations of the scalar potential $V$. 

\begin{figure}[!t]
\centerline{\includegraphics[width=9cm]{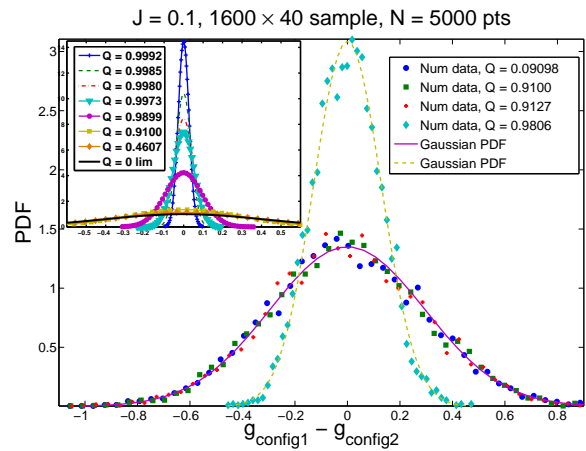}}
\caption{Probability Density Function of the difference 
$g[V,\{{\vec{S}}_{i}^{(1)}\}] - g[V,\{{\vec{S}}_{i}^{(2)}\}]$ as $V$ is varied. The result for various 
pairs of mean-field like spin states are shown, parametrized by the corresponding  overlap between these spin states. 
}
\label{fig:PDF_fctQ}
\end{figure}
 \begin{figure}[!t]
\centerline{\includegraphics[width=9cm]{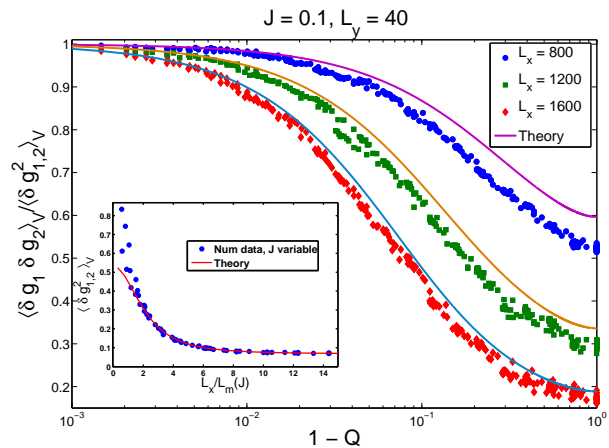}}
\caption{\label{fig:corr_Q}Conductance correlations as a function of spin configurations overlap for different longitudinal sizes, and normalized by their value for $Q_{12}=1$. 
In the inset, the conductance fluctuations ($Q_{12}=1$) are plotted as a function of $L_{x}/L_{m}$. 
}
\end{figure}

 We start by identifying the regime of weak localization. In this regime, for a given random spin 
configuration,  the variance of the above distribution of conductance 
$\langle (\delta g)^2\rangle = \langle \left( g[V, \{{\vec{S}}_{i}\}_{i}] - \langle g[V, \{{\vec{S}}_{i}\}_{i}] \rangle \right)^2 \rangle $
(where $\langle \rangle $ corresponds to an average over $V$)
is given in the 1D diffusive regime by 
\begin{multline}
 \langle (\delta g)^2\rangle 
=  
\frac{1}{4} F(0) + 
\frac{3}{4}F\left(\frac{2L_{x}}{\sqrt{3}L_{m}}\right) \\
+
  \frac{1}{4}F\left(\frac{2 L_{x}}{L_{m}} \right) + \frac{1}{4}F\left(\frac{\sqrt{2}L_{x}}{\sqrt{3}L_{m}}\right)
\label{eq:UCF}
\end{multline}
where we defined \cite{Pascaud:1998} 
$F(x)     = (6 + 6x^2 - 6\cosh(2x) + 3x\sinh x)/(x^4\sinh^2 x)$. 
 The amplitude of these fluctuations extrapolate  from $8/15$  for $L_{x}\ll  L_{m}$ 
 (orthogonal class with spin-degenerate states) to $1/15$ for $L_{x}\gg L_{m}$ 
 (unitary class with double number of modes).  The magnetic dephasing length $L_{m}$ 
 depends in particular on the strength of magnetic disorder $J$. It is numerically 
 determined through the use of formula (\ref{eq:UCF}). 
  The inset of Fig.  \ref{fig:corr_Q} shows the excellent agreement between the 
 numerical data for this conductance fluctuations and weak localization formula (\ref{eq:UCF}) plotted as a function of 
 $x=L_{x}/L_{m}(J)$. 

   Having determined the weak localization regime, we now turn to the study of 
 correlations of the conductance between two different spin configurations 
$\{\vec{S}_i^{(1)}\}_{i}$ and $\{\vec{S}_i^{(2)}\}_{i}$. 
We consider the distribution as $V$ is 
varied of the difference  
$ g\left[ V,\{\vec{S}_i^{(1)}\}\right] -   g \left[V,\{\vec{S}_i^{(2)}\}\right]$.  This distribution has naturally  zero mean, 
and
its variance encodes statistical correlations between the two conductances 
$\left<  (\delta g_1 -  \delta g_2)^2 \right> = 2 ( \left<  (\delta g_{1,2})^2 \right> - \langle \delta g_1 \delta g_2 \rangle ) $
  where $ g_{\alpha} =  g\left[ V,\{\vec{S}_i^{(\alpha)}\}\right]$. 
 Similarly to the variance (\ref{eq:UCF}), they are parametrized by four dephasing lengths  
 \cite{Altshuler:1985,Carpentier:2008,Fedorenko:2009} as follows 
\begin{multline}
\langle\delta g_1\delta g_2\rangle  = 
\frac{1}{4}F\left( L_{x}/L_{m}^{D,S} \right) + \frac{3}{4}F\left(L_{x}/L_{m}^{D,T} \right)
 \\
 + \frac{1}{4}F\left(L_{x}/L_{m}^{C,S}\right) + \frac{3}{4}F\left(L_{x}/L_{m}^{C,T}\right)
\label{equ:corr_fct_Q}
\end{multline}
 with $F(y)$ given in (\ref{eq:UCF}) and the $L_{m}^{C/D,S/T}$ correspond to magnetic dephasing 
 lengths for the Singlet/Triplet components of Diffuson/Cooperon contributions built between spin configurations 
 $1$ and $2$. We will study these conductance correlations for different types of correlations between 
 the spin configurations.

\paragraph{Mean-Field like excitations.} 
First, we consider spin states with no spatial correlations 
between them. These configurations are generated as follows : we start from a configuration $1$ where the orientations of 
spins are chosen randomly and independently from each other. From this first state, we generate other configurations by 
regenerating with a probability $p$ the orientations of each spin.  In this case, the overlap (\ref{equ:overlap}) is an adequate measure 
of the correlation between these states. In practice with this method we generated spin states with mutual overlap $Q_{12}$ from $10^{-3}$ to $1$.    For these spin configurations, we find a very good agreement with analytical studies\cite{Carpentier:2008} : the correlation between the conductances  is 
entirely parametrized by their overlap $Q_{12}$. 
In Figure \ref{fig:PDF_fctQ} we plot  the probability density function (PDF) of the difference $g[V,\{{\vec{S}}_{i}^{(1)}\}] - g[V,\{{\vec{S}}_{i}^{(2)}\}]$ as $V$ is varied. 
The PDF for three different pairs of spin states with the 
same overlap $Q_{12} \simeq 0.91$ (dots, squares and triangles) are identical with each other 
and different from the PDF for a pair with $Q_{12} = 0.98$. These PDF are found to be reasonably well  Gaussian, parametrized solely by the above second cumulant. 

\begin{figure}[!t]
\centerline{\includegraphics[width=9cm]{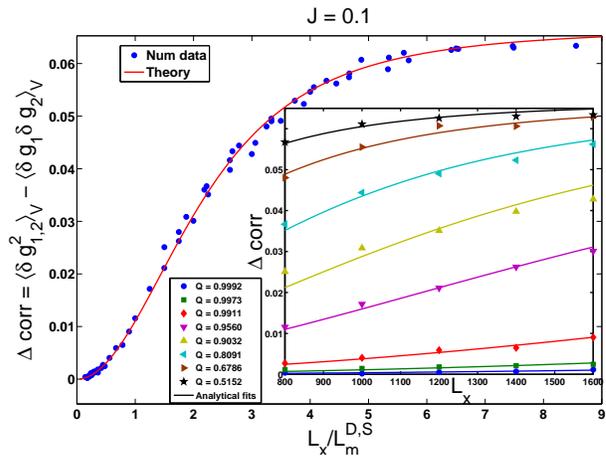}}
\caption{\label{fig:fitLmDSJ0p1}Comparison between the correlation between conductances 
$\sigma = 2 ( \langle (\delta g)^2\rangle_{V} - \langle\delta g_1\delta g_2\rangle_{V} )$ 
and the so-called Diffuson Singlet contribution  $F(0) - F(L_x/L_m^{D,S})$. 
 This contribution is found to be dominant in the region $Q_{12}\simeq 1$. 
In the inset we plot this $\sigma$ function for different value of overlap $Q_{12}$ and for $J = 0.1$. Plain lines are theoretical fits allowing to determine $L_m^{D,S}(Q_{12},J)$.  System size is $40 \times 1600$. 
}
\end{figure}

Then we compare the behaviour of this second cumulant with eq.(\ref{equ:corr_fct_Q}), using the analytical expressions 
to order $J^2$  for the dephasing lengths \cite{Carpentier:2008} :  
 $L_{m}^{D/C,S}= L_{m}/\sqrt{1\mp Q_{12}}$ and $L_{m}^{D/C,T}= L_{m}/\sqrt{1\pm Q_{12}/3}$. 
   We find a reasonable agreement between this prediction and numerical 
 results, as shown in Fig.~\ref{fig:corr_Q}.  Note that this comparison is done without 
 any free parameter as the magnetic dephasing length $L_{m}$ was determined from 
 $\langle (\delta g)^2\rangle $ (see Fig.~\ref{fig:fitLmDSJ0p1})
The behaviour of this variance also explains the high sensitivity of  $P( g_1 -  g_2 \simeq 0)$ on 
small departures from $Q_{12}=1$ as shown in the inset of Fig.~\ref{fig:PDF_fctQ}. 
Indeed, the probability of similar conductances reads $P(0) = 1/\sqrt{2\pi \sigma}$ with 
$\sigma = 2 ( \langle (\delta g)^2\rangle_{V} - \langle\delta g_1\delta g_2\rangle_{V} )$.

\begin{figure}[!t]
\centerline{
\includegraphics[width=8.5cm]{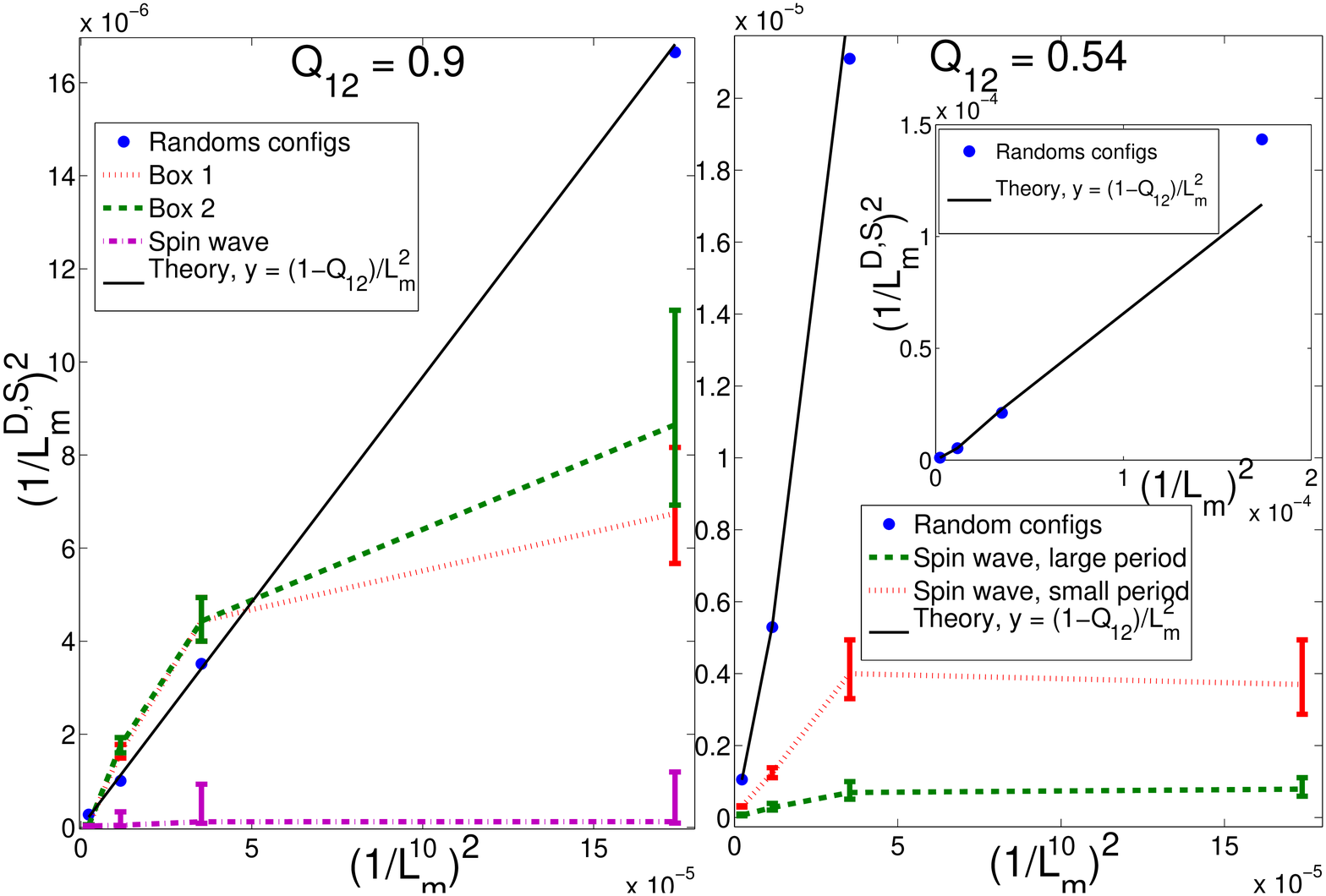}
}
\caption{\label{fig:invLmDSinvLm}Left:  evolution of the Diffuson Singlet dephasing rate of $1/(L_m^{D,S})^2$ 
as a function of the electronic dephasing rate $1/(L_m^2)$ for mean-field like, spin wave (dash dot), and boxed excitations. 
The overlap is $0.9$ for all of them, and the size is $40 \times 1600$.  
 The linear dependance corresponds to the analytical expression $1/(L_m^{D,S})^2=(1-Q_{12})/(L_m^2)$, valid in the absence 
 of spatial correlations between the spin states. These results show the clear departure from this behavior for 
 strong spatial correlations, and the absence of any effective overlap. 
 Right:  same evolution for random configurations and for spin wave with two different spatial periods corresponding to the same overlap ($Q=0.54$). The stronger correlation (longer period) corresponds to the larger deviation from the linear law. 
}
\end{figure}

From the above analytical expressions for $L_{m}^{C/D,S/T}$ for overlaps $Q_{12}\simeq 1$ the only diverging magnetic length 
is found to be $L_m^{D,S}$. Thus  in this regime the expression (\ref{equ:corr_fct_Q}) is dominated by the corresponding contribution \cite{Fedorenko:2009}. This allows for a direct determination of $L_m^{D,S}$ from the $L_x-$dependance of $\sigma$ 
in the region $Q_{12}\simeq 1$, as shown in Fig.~\ref{fig:fitLmDSJ0p1}. 
  We find an excellent agreement between the corresponding dephasing rate $1/(L_{m}^{D,S})^2$   and its perturbative analytical expression   $(1-Q_{12})/(L_{m})^2$ as shown on Fig.~\ref{fig:invLmDSinvLm}.

\paragraph{Correlated excitations.} 
We now consider the influence of  spatial correlation between two spin configurations. Generation of correlated spin 
configurations is obtained as follows:  from an initial spin configuration  we 
generate a configuration $n$  
by reversing spins preferably inside a box of size $L_{x}^{(n)}\times L_{y}$
in the middle of the sample. 
We consider boxes of increases length $L_{x}^{(n)} = n L_{x}^{(1)}$, such that 
all states have the same overlap $Q$ with the initial configuration, but their spatial correlations with this initial configuration decreases with $n$ (the larger the box, 
the smaller the probability that a given spin inside the box is modified). 
 We also generated  spin wave like excitations: from the same initial spin configuration,  each spin is rotated 
 by $\delta\phi (x,y) = x\delta\phi_0$ around the $z-$axis (axis perpendicular to the planar sample). 
 $\delta\phi_0$ determines the period of the spin wave, and thus the overlap between both configurations.  
  For the different pairs of correlated spin configurations, we repeat the previous analysis of conductance correlations. 
  In particular, we determine the Diffuson Singlet length $L_m^{D,S}$ for different values of magnetic disorder
   amplitude $J$ in the region $Q_{12}\simeq 1$. The result is shown in the left part of Fig. \ref{fig:invLmDSinvLm} for the overlap $Q_{12}=0.9$.
   We find clear deviation  from the behavior 
  $1/(L_{m}^{D,S})^2 = (1-Q_{12})/(L_{m})^2$ which is valid in the absence of spatial correlations (Fig.~\ref{fig:fitLmDSJ0p1}). 
  The deviation from this linear behavior is largest for the strongest spatial correlations between spin states, {\it i.e} for the smallest box excitations ($n=1$) and the spin wave excitations. We also 
  consider two spin wave excitations with different period but the same overlap with an initial spin state (Fig.~\ref{fig:invLmDSinvLm}). 
 Here again, the resulting Diffuson Singlet magnetic length is smaller for 
 the strongest spatial correlation between the two spin states, corresponding to 
 the largest period.    
  These results demonstrate the sensitivity of the magnetic dephasing length 
 $L_{m}^{D,S}$ on spatial correlations between the magnetic disorder configurations, 
 and thus the influence of the 
  geometry of random spin excitations on the associated correlation of conductances. 
  Note that these results of Fig.~\ref{fig:invLmDSinvLm} can in principle be tested experimentally by varying 
 the density of magnetic impurities while working at fixed $T/T_{SG}$, allowing for an unprecedented test of 
 {\it e.g} the nature of excitations in a spin glass state.

In this letter we presented a numerical Landauer analysis of transport in a mesoscopic metallic wire in the presence of frozen 
magnetic impurities. We have found that statistical properties of conductance correlations between two mean-field like spin configurations 
depend only on the corresponding spin overlap, in agreement with theoretical analysis. These  results open the route to 
direct spin state correlations in mesoscopic spin glasses. We have also shown the crucial importance of spatial correlations between spin configurations in the electronic dephasing process. Studying these correlations along the lines of 
Fig. \ref{fig:invLmDSinvLm} could be experimentally achieved by varying the electronic density in diluted 
magnetic semiconductors. Unfortunately this would also modify the couplings between the impurity spins, and hence 
the spin configuration. A more promising route consists in exploring other multi-terminal geometries along the lines of \cite{deVegvar:1991}. 
\\
\indent 
We thank T. Capron for a much informative discussion concerning possible experimental test of our results.  
This work was supported by the ANR grants QuSpins and Mesoglass. All numerical calculations 
were performed on the computing facilities of the ENS-Lyon calculation center (PSMN).


\end{document}